\documentclass[prb,twocolumn,groupedaddress,showpacs]{revtex4}

\usepackage{graphicx,amsmath}
\newcommand{\beq}{\begin{equation}}
\newcommand{\eeq}{\end{equation}}
 
\newcommand{\ent}{{\mathcal S}}
\newcommand{\SK}{S_\mathrm{Kelvin}}

\begin{document}

\title{The  Kelvin  Formula for Thermopower}

\author{Michael R. Peterson}
\affiliation{Condensed Matter Theory Center, Department of Physics, University of Maryland, College Park, MD 20742, USA}
\author{B. Sriram Shastry}
\affiliation{Physics Department, University of California,  Santa Cruz, CA  95064, USA}

\begin{abstract}
Thermoelectrics are important in physics, engineering, and material
science due to their useful applications and inherent
theoretical difficulty. Recent experimental interest has shifted to  strongly correlated
materials, where the calculations become particularly difficult. Here we reexamine the framework for calculating the thermopower, inspired by ideas of Lord
Kelvin from  1854. We  find an approximate but concise expression, 
which we term as {\em the Kelvin formula} for the the Seebeck coefficient.
According to this formula,  the Seebeck coefficient is given 
as the particle  number $N$  derivative of the entropy $\ent$, at 
constant volume $V$ and temperature $T$,
$S_{\text{Kelvin}}=\frac{1}{q_e}\left\{ \frac{\partial {\ent
}}{\partial N} \right\}_{V,T}$.  This formula is
shown to be competitive compared to other approximations in various
contexts including strongly correlated systems. We finally  connect to a
recent thermopower calculation for non-Abelian fractional quantum Hall
states, where we point out that  {\em the Kelvin formula is exact}.
\end{abstract}

\date{\today}


\maketitle

\section{Introduction}

A complete understanding of thermoelectric
effects is  important in the physical sciences 
where wide ranging applications utilize
materials with large thermoelectric power
$S$ (Seebeck coefficient).  Thermoelectrics of strongly correlated
materials are of fundamental interest 
since they present an important and challenging problem.  Recent experiments
have revealed that some materials, such as sodium cobalt oxide
Na$_x$CoO$_2$ (NCO), possess unusually large
thermopower~\cite{nco-terasaki-wang-ong}, due in part  to
strong electron interactions~\cite{mrp-joh-bss-prl-prbs}.
Frustrated systems~\cite{frustration}, like NCO, might produce further
surprises in enhanced thermopower in some
situations~\cite{bss-prb,mrp-joh-bss-prl-prbs}.  In addition, emerging 
work~\cite{yang-halperin} from the fractional quantum Hall
effect (FQHE) is revitalizing thermopower as a tool to
investigative the topological non-Abelian quasiparticles~\cite{mr-pf}
thought to exist at filling factor 5/2~\cite{willett}.

Here we present the Kelvin formula for thermopower,
 $S_\mathrm{Kelvin}$. This  is a  formula inspired by Lord Kelvin's thermodynamic
 treatment of this variable in 1854~\cite{kelvin}. It  is
 found by reconsidering the sequence of taking the thermodynamic and
 uniform limits, and is a valuable approximation to the exact, but
computationally intractable result, obtained via Onsager and Kubo's
 treatments~\cite{onsager,kubo}.

For strongly correlated systems, such as the $t$-$J$ model,
$S_\mathrm{Kelvin}$ is found to possess an accuracy between the rather
coarse  Mott-Heikes formulation, and a better argued  high frequency limit
formulation due to Shastry~\cite{bss-prb} and studied in
Refs.~\onlinecite{mrp-joh-bss-prl-prbs} and ~\onlinecite{bss-review}.  For
intermediate couplings, such as the Hubbard model, we argue that
$S_\mathrm{Kelvin}$  provides
one of the best available approximations, it  is better than the high frequency limit.
  In certain dissipation-less
situations, such as the FQHE, $S_\mathrm{Kelvin}$ is exact, thereby
providing an elegant and simple derivation for the thermopower formula
used in Ref.~\onlinecite{yang-halperin} (derived originally in
Refs.~\onlinecite{obraztsov-cooper}.)

$\SK$ is obtained by completing Shastry's argument~\cite{bss-review}
for the ``absolute thermopower'', i.e. $S$ of an isolated system.
Kelvin originally studied~\cite{kelvin} this object using the then available
techniques, later he and others emphasized relative thermopower between two
materials. Let us revert to the absolute thermopower as a starting
point and imagine a long isolated cylinder of material of length $\L$
subject to a time dependent electric field $- \nabla \Phi$ and
temperature gradient $\nabla T$.  $- \nabla \Phi$ couples to the
dipole moment and $\nabla T$ couples to the moment of the energy
density (cf.  Luttinger~\cite{luttinger}).  These fields individually
generate a dipole moment linear in the fields to lowest order, and the
condition for the cancellation of the two contributions, i.e., the
zero dipole moment (or zero current) condition, leads to the
thermopower $S$ for a finite system size $\L$ at finite frequencies
$\omega$ as $S(\L,\omega)= \frac{ \nabla \Phi}{\nabla T}\{\L, \omega
\}$.

 The thermodynamic limit, $\L \rightarrow \infty$, and the static
 limit, $\omega \rightarrow 0$, must both be taken, as known from
 Onsager~\cite{onsager} and others~\cite{edwards,luttinger}.  Kubo's
 exact formulas obtain in {\em the fast or  transport limit}, where $\L \to \infty$
 before $\omega \to 0$.  Taking the static limit $\omega \to 0$ before
 $\L \to \infty$ leads to  {\em the slow }, where Kelvin's approximate formula arises and
 is expressible solely in terms of equilibrium  thermodynamic variables.

 We transcribe this discussion to a more convenient periodic system,
  by trading the length scale $\L$ for a wave vector $q_x= 2 \pi/\L$
  and the $\L \rightarrow \infty$ limit by the uniform limit $q_x
  \rightarrow 0$.  The slow limit corresponds to $\lim{\{q_x
  \rightarrow 0, \; \omega \rightarrow 0 \} }$, and the fast limit
  corresponds to $\lim{\{ \omega \rightarrow 0,\; q_x \rightarrow 0
  \}}$.  
  The thermopower measures the induced thermoelectric voltage due to a 
  temperature gradient and, as such, a useful and general formula for thermopower is
  given by the ratio between the thermoelectrical and 
  electrical conductivities~\cite{bss-review},
\begin{eqnarray}
 S(q_x,\omega) & = & \frac{ \chi_{\rho(q_x), \hat K(-q_x)}(\omega)}{T
 \ \chi_{\rho(q_x), \rho(-q_x)}(\omega)} \;,
 \label{kelvin_1}
\end{eqnarray}
where
\begin{eqnarray}
\chi_{\hat A,\hat B}(\omega) &=& i \int_0^\infty dt e^{( i\omega
-0^+)t} \langle[ \hat A(t),\hat B(0)] \rangle \\
&=&\sum_{n,m}\frac{p_n-p_m}{\varepsilon_m-\varepsilon_n+\omega}\langle n|\hat{A}|m\rangle
\langle m|\hat{B}|n\rangle
\label{suscep}
\end{eqnarray}
 is the susceptibility
of any two operators $\hat A$ and $\hat B$, where $\rho $, $\hat K=
\hat H- \mu \hat{N}$, and $\hat{J}_x$ are the charge density, the
(grand) Hamiltonian, and the charge current operator, respectively, at
finite wave vectors; $\hat H$, $\mu$, and $\hat{N}$ are the
Hamiltonian, the chemical potential, and the total number operator,
respectively.  The susceptibility written in 
 Eq.~(\ref{suscep}) is the Lehmann representation  (where $p_n=\exp(-\beta\varepsilon_n)/Z$ is the probability of the quantum state $|n\rangle$ with energy $\varepsilon_n$ and $Z$ is the partition function and $\beta=1/k_\mathrm{B}T$
with $k_\mathrm{B}$ the Boltzmann constant) which 
 we find useful below.  With Eq.~(\ref{kelvin_1}), we can take different limits and
obtain various interesting formulas.

\section{Thermopower formulae}

\subsection{The Kubo formula}
Taking the fast limit and using the continuity
equations to pass from densities to current operators,
Eq.~(\ref{kelvin_1}) gives the exact Kubo result~\cite{kubo}
\begin{eqnarray}
S_{\mathrm{Kubo}} =\frac{1}{T} \frac{\int_0^\infty dt \int_0^\beta d
  \tau \langle \hat{J}^E_x( t - i \tau) \hat{J}_x(0) \rangle
}{\int_0^\infty dt \int_0^\beta d \tau \langle \hat{J}_x( t - i \tau)
  \hat{J}_x(0) \rangle} -\frac{\mu(T)}{q_eT} \label{kubo_1},
\end{eqnarray}
where $q_e$ is the charge of the carriers and $\hat{J}^E$ the energy
current.

\subsection{The Mott-Heikes formula} 
For narrow band systems, such as
NCO, high $T_C$ superconductors, or heavy fermion systems, the
so-called Mott-Heikes (MH) approximation introduced by Heikes
(popularized by Mott~\cite{beni-heikes}) is written $S_\mathrm{MH} =
(\mu_0 - \mu(T))/q_e T$, where $\mu_0\equiv\mu(T=0)$.  $S_\mathrm{MH}$
is obtained by rather drastically replacing the first part of
Eq.~(\ref{kubo_1}) by the zero temperature chemical potential $\mu_0$
to make the theory sensibly behaved as $T \to 0$. From thermodynamics,
we know that $ -\frac{\mu(T)}{T}= (\frac{\partial {\ent}}{\partial
N})_{E,V}$, and, hence, $S_\mathrm{MH}$ relates thermopower to the
partial derivative of entropy $\ent$ with particle number $N$, {\em
at a fixed energy $E$ and volume $V$}.  We see below that $\SK$ is
similar, but with more natural ``held'' variables, namely $T$ and $V$.

\subsection{High frequency formula}
From Eq.~(\ref{kelvin_1}), we can
 make a high frequency approximation, where $\omega \gg \omega_c$
 ($\omega_c$ representing all finite characteristic energy scales),
 leading to the object $S^*$.  The formal expression and evaluation
 for $S^*$ are discussed elsewhere~\cite{bss-review} and we only quote
 the results.  We have argued that $S^*$ is the best possible
 approximation to the exact Kubo formula for strongly correlated
 systems~\cite{mrp-joh-bss-prl-prbs} such as the $t$-$J$ model, since
 the high frequency limit respects the single occupancy constraint and
 is closer to the DC limit than initially expected.  It is not
 specifically suited for Hubbard type models, since the high frequency
 limit assumes $\omega \gg U$, and cannot capture the physics of
 correlations effectively~\cite{bss-review}.  We will see that $\SK$
 steps into this breach, and provides a very useful alternative for
 Hubbard type models~\cite{louis}.

\subsection{The Kelvin formula}
To obtain an approximate
thermodynamical expression, we consider the slow limit of
Eq.~(\ref{kelvin_1}).  $S$ is among the few objects (along with
Hall constant and Lorentz number) where this process gives finite
and approximate results, unlike the electrical conductivity where
the slow limit gives meaningless results~\cite{bss-review}.  
\begin{widetext}
This
limit is identified with Kelvin since he essentially took the
equilibrium limit of an interacting gas of particles.  The slow limit
($q_x\rightarrow 0$, $\omega\rightarrow 0$) is easiest to compute
starting from Eq.~(\ref{kelvin_1})
\begin{eqnarray}
S_\mathrm{Kelvin} = \lim_{q_x \rightarrow 0} \frac{ \chi_{ \rho(q_x),
\hat K(-q_x)}(0)}{T \ \chi_{ \rho(q_x), \rho(-q_x)}(0)}.
\label{kelvin_0}
\end{eqnarray}
To simplify we first consider the numerator of Eq.~(\ref{kelvin_0}) which we rewrite by 
first using the Lehmann representation and then taking the $q_x\rightarrow0$ limit.  
Note that 
$\hat{\rho}(q_x)$ tends to a conserved quantity $q_e N$ and cannot mix states of 
different energy so $\varepsilon_m\rightarrow\varepsilon_n$.  Thus, 
\begin{eqnarray}
\lim_{q_x \rightarrow 0} \chi_{ \rho(q_x),\hat K(-q_x)}(0)
&=&\lim_{q_x\rightarrow0}
\sum_{n,m}\frac{p_n-p_m}{\varepsilon_m-\varepsilon_n}\langle n|\hat{\rho}(q_x)|m\rangle
\langle m|\hat{K}(-q_x)|n\rangle\nonumber\\
&=&\lim_{\varepsilon_m\rightarrow\varepsilon_n}\sum_{n,m}\frac{p_n-p_m}{\varepsilon_m-\varepsilon_n}\langle n|\hat{\rho}(q_x)|m\rangle\langle m|\hat{K}(-q_x)|n\rangle\nonumber\\
&=&\lim_{\varepsilon_m\rightarrow\varepsilon_n}\sum_{n,m}p_n\frac{1-e^{\beta(\varepsilon_n-\varepsilon_m)}}{\varepsilon_m-\varepsilon_n}\langle n|\hat{\rho}(q_x)|m\rangle\langle m|\hat{K}(-q_x)|n\rangle\nonumber\\
&=&q_e\sum_{n,m}\beta p_n\delta_{\varepsilon_n,\varepsilon_m}\langle n|\hat{N}|m\rangle\langle m|\hat{K}|n\rangle\nonumber\\
&=&q_e\beta[\langle\hat{N}\hat{K}\rangle-\langle\hat{N}\rangle\langle\hat{K}\rangle]\nonumber\\
&=&q_e\left[ \frac{d}{d \mu}
\langle \hat H \rangle - \mu \frac{d}{d \mu} \langle \hat{N} \rangle
\right]\;.
\label{spht3}
\end{eqnarray}

 The derivative with respect to $\mu$ in Eq.~(\ref{spht3}) is
 within the grand canonical ensemble and performed with a fixed $V$
 and $T$. The denominator of Eq.~(\ref{kelvin_0}) is treated
 similarly yielding $q_e^2\beta[\langle\hat{N}^2\rangle-\langle\hat{N}\rangle^2]=q_e^2d\langle\hat{N}\rangle/d\mu$.  Combining it with Eq.~(\ref{spht3}), yields
\begin{eqnarray}
S_\mathrm{Kelvin} & = & \frac{1}{q_eT} \frac{\frac{d}{d \mu} \langle
\hat H \rangle - \mu \frac{d}{d \mu} \langle \hat{N} \rangle
}{\frac{d}{d \mu} \langle \hat{N} \rangle }\;.
\label{skelvin-original}
\end{eqnarray}

\end{widetext}

To further simplify Eq.~(\ref{skelvin-original}) we note a  
relation found in textbooks on thermodynamics 
in the grand canonical ensemble: $ \langle \hat H \rangle\equiv E=
\Omega + T \ent + \mu N$ ($\Omega$ the grand potential), so that
$(\frac{\partial E}{\partial \mu})_{T,V}= \mu (\frac{\partial N
}{\partial \mu})_{T,V}+ T (\frac{\partial \ent}{\partial \mu})_{T,V} $
and hence
\begin{eqnarray}
S_\mathrm{Kelvin}  &=&  \frac{1}{ q_e} \frac{(\frac{\partial
\ent}{\partial \mu})_{T,V}}{(\frac{\partial N }{\partial \mu})_{T,V} }\nonumber\\
\label{kelvin_formula_entropy1}
&=& \frac{1}{ q_e} \left(\frac{\partial \ent}{\partial N}\right)_{T,V}\\
\label{kelvin_formula_entropy2}
 &=& \frac{-1}{ q_e} \left(\frac{\partial \mu}{\partial T}\right)_{N,V}
\end{eqnarray} 
where we used, to go from the second equality to the last 
equality,  a Maxwell relation obtained with
$dF= - {\ent} dT - p dV + \mu dN$, and equating $\frac{\partial^2
F}{\partial T \partial N}=\frac{\partial^2 F}{\partial N \partial T}
$. We refer to the last two equivalent equations (Eqs.~(\ref{kelvin_formula_entropy1}) and~(\ref{kelvin_formula_entropy2}))
as the Kelvin formula for the thermopower.
This formula is unknown in the literature as far as we are aware.

Note that $S_\mathrm{MH}$ is
similar to $S_\mathrm{Kelvin}$. The distinction is that in $\SK$, the
number derivative of the entropy is taken at constant $T$ rather than
at constant $E$. Thus, in the low $T$ limit of a metal, where
$\mu(T)\propto T^2 $, they differ in the linear $T$ coefficient by a
significant factor of two.  We show below that for non-interacting
electrons, scattered by impurities, $\SK$ is closer to the exact
result than $S_\mathrm{MH}$.  Further, we see that the approximation
of exchanging the slow and fast limits has some justification in
dissipation-less systems, such as in the FQHE where
$\SK$ is identical to that found by several workers (see below).

\section{Applications of thermopower formulae}

\subsection{Free electrons}
To gain insight into the strengths and
weaknesses of the various thermopower formulations discussed above we
consider  non-interacting degenerate electrons treated
within the limit of elastic scattering at the Born level with an
energy momentum dependent relaxation time $\tau(p,\omega)$.  This is a
modestly dissipative system, but at such a simple level that the
Boltzmann-Bloch equation is an adequate description.  The solution for
$S$ is available in textbooks
and a useful benchmark for various approximations.  In the low
temperature limit~\cite{bss-review}, to $\mathcal{O}(T^3)$, 
\beq S_\mathrm{Mott} = T
\frac{ \pi^2 k_B^2 }{ 3 q_e} \frac{d}{d \mu} \ln \left[ \rho_0(\mu)
\langle (v^x_{p})^2 \tau(p,\mu)\rangle_{\mu} \right] |_{\mu
\rightarrow \mu_0}\; ,
\label{mott}
\eeq 
a formula often ascribed to Mott and $\rho_0(\mu)$ is the single
particle density of states per unit volume per spin. In this
non-interacting electron context, $\SK$ gives (to $\mathcal{O}(T^3)$), 
\beq S_{\mathrm{Kelvin}} = T \frac{ \pi^2
k_\mathrm{B}^2 }{ 3 q_e} \frac{d}{d \mu} \ln \left[ \rho_0(\mu)
\right]|_{\mu \rightarrow \mu_0} \;,
\label{kelvin-mott}
\eeq 
which differs from the exact answer 
(Eq.~(\ref{mott})) in the neglect of the relaxation time $\tau$
and particle velocity $v^x_p$ in the logarithm. $S_\mathrm{MH}$, to
the same order, gives 
\beq 
S_\mathrm{MH} = T \frac{ \pi^2
k_\mathrm{B}^2 }{ 6 q_e} \frac{d}{d \mu} \ln \left[ \rho_0(\mu)
\right]|_{\mu \rightarrow \mu_0}
\eeq 
which is off by an important
factor of two from $\SK$ (Eq.~(\ref{kelvin-mott})).  The
formulations (Mott-Heikes and Kelvin) would be identical if
$\mu\propto T$, which occurs if the system possesses a ground state
degeneracy, and in the classical regime. 
The high frequency approximation gives a better result than all these
and, again in the low temperature limit, to
$\mathcal{O}(T^3)$, 
\beq 
S^* = T \frac{ \pi^2 k_\mathrm{B}^2 }{ 3 q_e}
\frac{d}{d \mu} \ln \left[ \rho_0(\mu) \langle (v^x_{p})^2
\rangle_{\mu} \right]|_{\mu \rightarrow \mu_0}\;.
 \label{high-freq}
\eeq 
Other than the neglect of the energy derivative of $\tau$, this
is the same as the exact result.  Hence,
ranking the thermopower approximations for non-interacting electrons
we have, from worst to best, $S_\mathrm{MH}$, $S_\mathrm{Kelvin}$, and
$S^*$ with the exact result being $S_\mathrm{Mott}$.

\subsection{Hubbard model}
For intermediate coupling models, the
 relative rankings of the various approximations can be different. In
 particular, $\SK$ can be superior to $S^*$, since the effect of
 correlations is diluted in the latter by making the assumption of
 $\omega \gg U$, whereas $\SK$ retains $\omega \ll U$.  The sign of
 the true (i.e. transport) thermopower and the transport Hall constant
 are expected to flip as we approach half filling in the Hubbard or
 $t$-$J$ models due to the onset of correlations (carriers become
 holes measured from half filling rather than from a completely filled
 band). In the case of the $t$-$J$ model, the high frequency Hall
 constant $R_H^*$ and $S^*$ do display this
 behavior~\cite{bss-review}. However, for the Hubbard model,
 $R_H^*$ and $S^*$ do not display a sign
 change~\cite{assad,louis}. $\SK$ on the other hand, does appear to show the expected
 change in sign~\cite{louis,pruschke}.  Further discussion concerning
 the relative merits of $\SK$ and $S^*$ will be reported
 later~\cite{louis}.

\subsection{NCO and the $t$-$J$ model}
\begin{figure}[t]
\begin{tabular}{cc}
\includegraphics[width=8.0cm,angle=0]{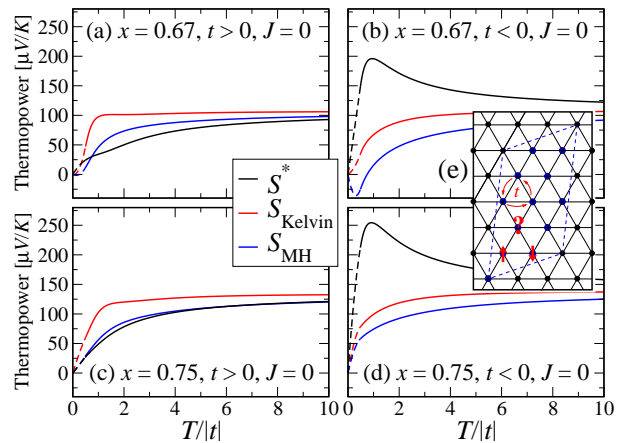}
\end{tabular}
\caption{Thermopower vs. $T$ for the $t$-$J$ model (with $J=0$)
corresponding to NCO in the Curie-Weiss metallic phase near
$x\sim0.7$. (a) and (c) correspond to $x=0.67$ and $x=0.75$ for the
NCO system ($t>0$) while in (b) and (d) the sign of the hopping has
been switched to investigate the enhancement expected for frustrated
systems.  The black, red (light gray), and blue (dark gray) lines are $S^*$,
$S_\mathrm{Kelvin}$, and $S_\mathrm{MH}$.  Finite size effects at low
$T$ are treated in the spirit described
previously~\cite{mrp-joh-bss-prl-prbs}.  At each $x$, for $T$ below
an appropriately chosen cutoff temperature $T_0=0.5|t|$, the thermopower is fit to $S(T)\to aT+bT^2$
where $a$ and $b$ are obtained from the computed $S(T_0)$ and
$S'(T_0)$ providing a sensible extrapolation to low $T$ and plotted as
dashed lines.  The inset figure (e) depicts the 12-site unit cell.}
\label{plot-ss}
\end{figure}

To show the usefulness of
$\SK$, we apply it to NCO since (i) we have previously
investigated~\cite{mrp-joh-bss-prl-prbs} this system while
benchmarking $S^*$, (ii) the system is intrinsically
interesting~\cite{nco-terasaki-wang-ong}, and (iii) we can compare
different thermopower formulations on equal footing.  As
discussed~\cite{mrp-joh-bss-prl-prbs}, the action in NCO takes place
primarily in the cobalt oxide planes where $d$-shell spin-1/2
electrons live on a triangular lattice and these strongly interacting
2D electrons can be modeled with the $t$-$J$ model.  Hence, we exactly
diagonalize the $t$-$J$ model on a $\L=12$ site two-dimensional
triangular lattice with periodic boundary conditions
(cf. Fig.~\ref{plot-ss}e).  Note that we only show results for the
$t$-$J$ model with zero super-exchange interaction ($J=0$), as the
results only weakly depend on $J$. To map the $t$-$J$ model to NCO we
follow Refs.~\onlinecite{mrp-joh-bss-prl-prbs}
and~\onlinecite{bss-prb} and give results as a function of electron
doping $x=|1-n|$ away from half filling ($n$ is electron number
density).

\begin{figure}[t]
\begin{tabular}{cc}
\includegraphics[width=3.5cm,angle=-90]{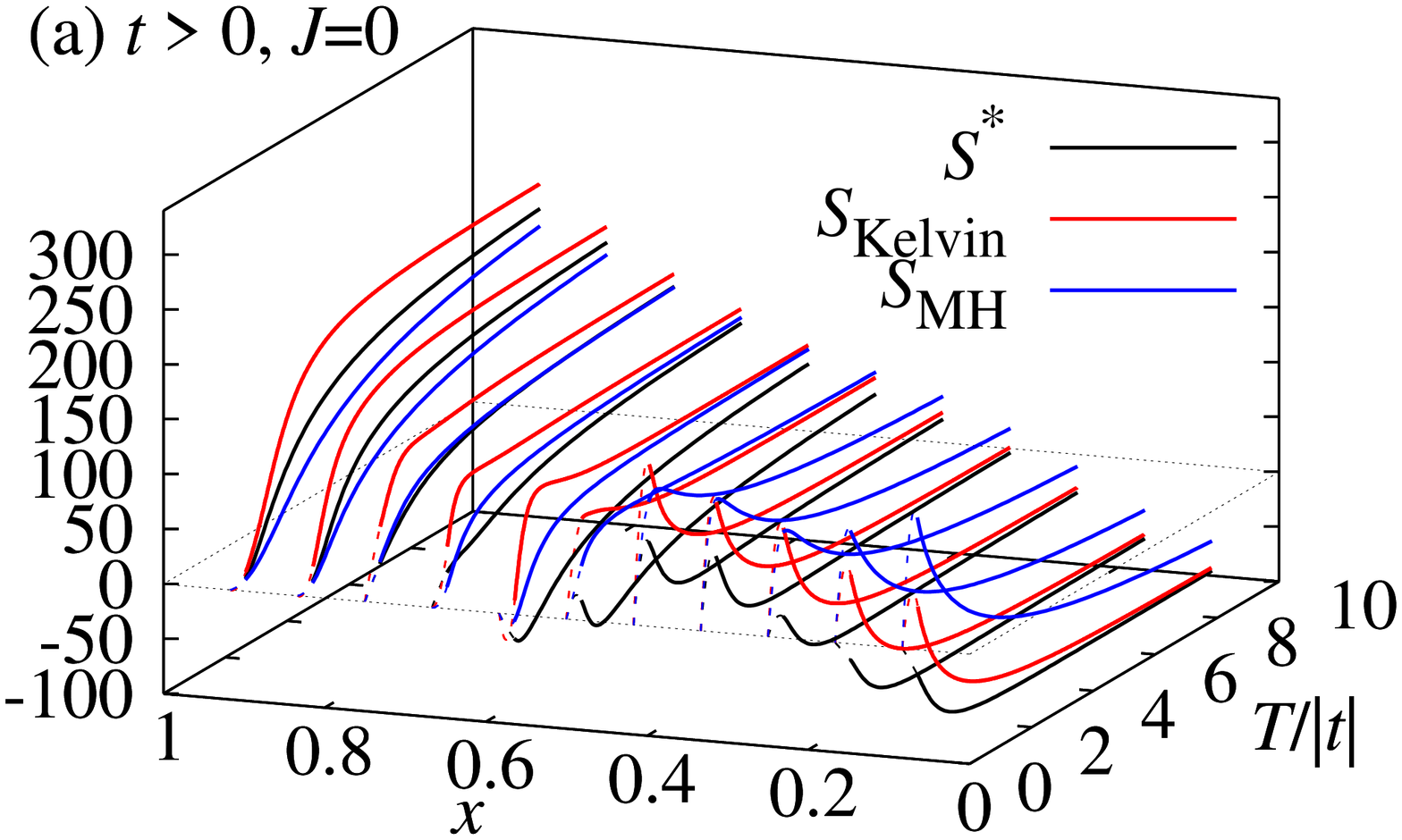}\\
\includegraphics[width=3.5cm,angle=-90]{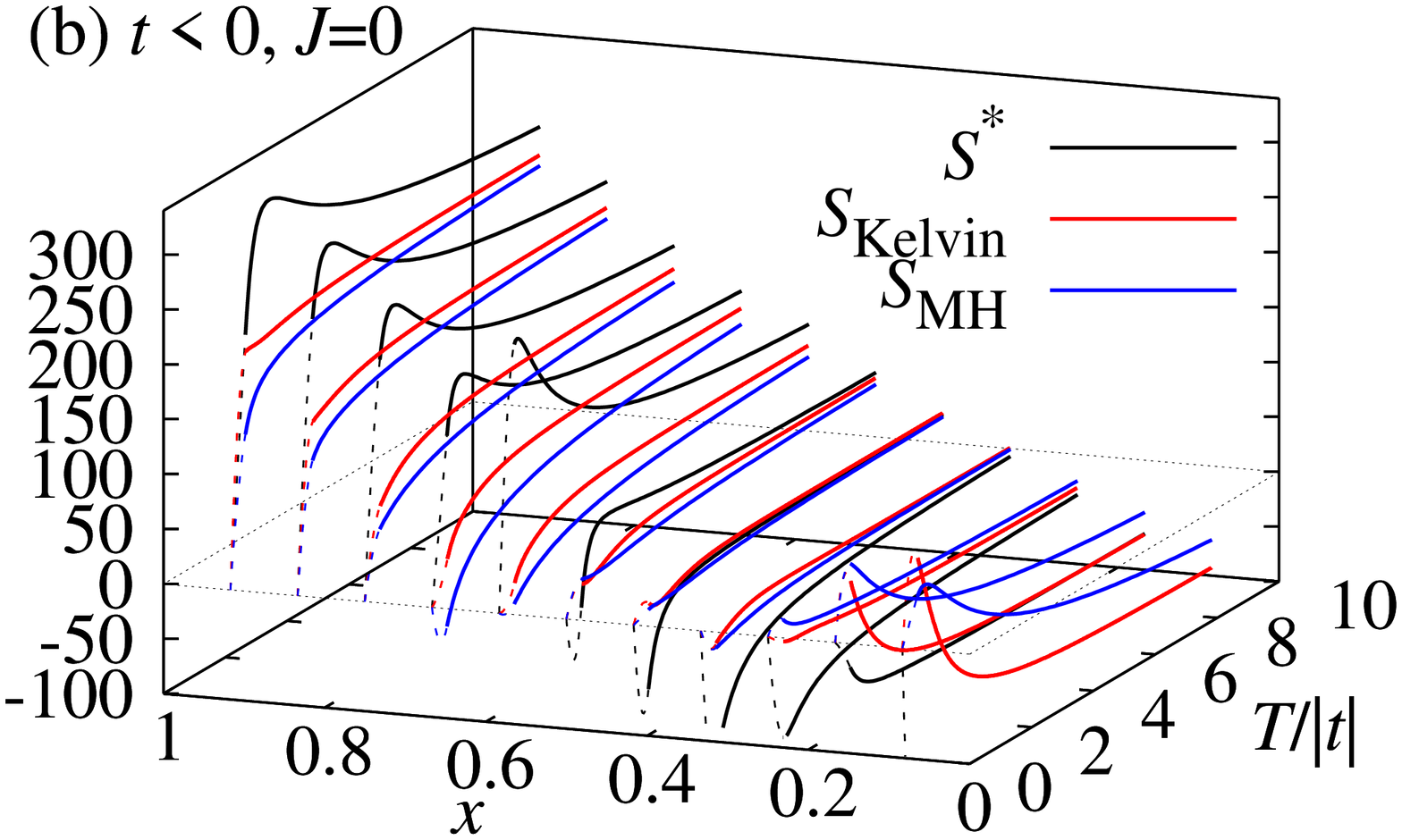}
\end{tabular}
\caption{The thermopower vs. $T$ and 
$x$ for the $t$-$J$ model (with $J=0$) for $t>0$ (a) and
$t<0$ (b).  Note that $x\sim0.7$ corresponds to the Curie-Weiss
metallic phase of NCO, cf. Fig.~\ref{plot-ss}.  The line-type and
color-coding is the same as in Fig.~\ref{plot-ss}.  Note that for dopings
below 0.5 it is not clear whether the $t$-$J$ model adequately
describes the physics of NCO.}
 \label{plot-s}
\end{figure}

$S^*$ adequately describes the physics of NCO for $x>0.5$ and, in
particular the so-called Curie-Weiss metallic phase~\cite{mrp-joh-bss-prl-prbs} near
$x\sim0.7$.  The subject of this work,
however, is $\SK$.  We see in Figs.~\ref{plot-ss}a and c and
~\ref{plot-s}a, similar to $S_\mathrm{MH}$, $\SK$ does a good job
capturing the physics with minimal computational effort.  However,
$S_\mathrm{Kelvin}$ does seem to overestimate the thermopower for
intermediate temperatures and high dopings as compared to
$S_\mathrm{MH}$.  Near $x\approx 0.7$, $\SK$ and $S_\mathrm{MH}$ are
similar but as $x$ is decreased the two formulae diverge and for low
dopings, $S_\mathrm{Kelvin}$ better captures the physics as it is
closer to the more accurate high frequency limit $S^*$.

An interesting property of the triangular lattice underlying the
physics of NCO is its geometrical frustration~\cite{frustration},
cf. inset Fig.~\ref{plot-ss}e.  It was
predicted~\cite{bss-prb,mrp-joh-bss-prl-prbs} that if the sign of the
hopping amplitude were flipped to $t<0$ the thermopower would be
enhanced at low to intermediate $T$.  We have considered this
situation in Figs.~\ref{plot-ss}b and d and~\ref{plot-s}b.  Since the
thermopower enhancement for $t<0$ compared to $t>0$ is largely a
consequence of electron-electron interaction it is important to
determine whether this effect is captured by $S_\mathrm{Kelvin}$.  We
see this enhancement is captured to some extent by $S_\mathrm{Kelvin}$
and $\SK$ is better than $S_\mathrm{MH}$ in the large doping region
where the enhancement is the greatest, but is missing some of the
electron-electron physics at very low $T$ that is captured by $S^*$
(as is $S_\mathrm{MH}$).

\subsection{FQHE at $\nu=5/2$}
We now discuss how $S_\mathrm{Kelvin}$
is applied to dissipation-less systems such as the FQHE where
thermopower can be used as a possible non-Abelian quasiparticle
detector~\cite{yang-halperin}.

For a weakly disordered electron system (from Eq.~(\ref{mott})
and~(\ref{kelvin-mott})) $\SK$ essentially gives the dissipation-less
thermopower where particle velocities are further approximated.  If
the system is dissipation-less and the particle velocities are also
energy independent, such as the FQHE, then we expect
$S_\mathrm{Kelvin}$ is exact.  An expression for the thermopower in a
2D electron system in the presence of a perpendicular magnetic field
(the FQHE system) has been
derived~\cite{obraztsov-cooper,yang-halperin}, assuming zero
impurities, as $\frac{\ent}{q_e N}$ (Eq.~(6) in
Ref.~\onlinecite{yang-halperin}).  Yang and Halperin
show~\cite{yang-halperin} that $\ent\sim k_B N\log (d)$ where $d>1$
is the quantum dimension of the quasiparticles for the FQHE at
$\nu=5/2$ (provided they are non-Abelian). Thus, a non-zero entropy
linear in $N$ is obtained.  From Eq.~(\ref{kelvin_formula_entropy1}),
we see that the thermopower is the derivative of the entropy with
respect to the number of particles at constant $T$ and $V$.  When
entropy is linear in particle number, as in non-Abelian FQHE states,
$\partial\ent/\partial N\rightarrow \ent/N$ and the formulas are
identical.  Our derivation provides a simple and straightforward
insight into the formula given previously~\cite{yang-halperin}.

\subsection{High Temperature Superconductors}
Before concluding, we
point out an intriguing application of $\SK$ for high $T_c$
superconductors. For different families of high $T_c$ compounds, a
universal curve of the thermopower, at $T=290 K$, as a function of
hole concentration $p\sim 1-n$ has been observed~\cite{tallon-honma}.
The thermopower, in all families, vanishes near optimal doping ($p\sim
0.16$) starting out positive at small $p$. Phillips \emph{et
al}.~\cite{phillips} appeal to the atomic limit of $S_\mathrm{MH}$ as
an explanation.  Viewing this data ~\cite{tallon-honma} more
generally, through the prism of $\SK$
(Eq.~({\ref{kelvin_formula_entropy1})) we conclude that the optimal
filling, i.e., maximum $T_c$, additionally corresponds to a {\em local
maximum of the electronic entropy as a function of filling}. This
conclusion is powerful, since we avoided the difficult issue of
calculating either thermopower or entropy, merely using the link
between them provided by $\SK$.

\section{Conclusion}
It is clear that $S_\mathrm{MH}$, which has served
as a virtual workhorse for years, has a new competitor in
$S_\mathrm{Kelvin}$. This simple minded approximation can be written
in closed form and in many difficult regimes, where the exact
Kubo-Onsager expressions are not useful, and $\SK$ provides an
excellent guide to the physics of the system.

We acknowledge support from DOE-BES Grant No. DE-FG02-06ER46319 at
UCSC, and MRP acknowledges support from LPS-CMTC at University of
Maryland and Microsoft Project Q.

\end{document}